\documentclass{czjphyse}
\usepackage{epsfig}  
\newcommand{\be}{\begin{eqnarray}}
\newcommand{\ba}{\begin{array}}
\newcommand{\ea}{\end{array}}
\newcommand{\ee}{\end{eqnarray}}

\newcommand{\dslash}{\partial \hskip -0.5em /}

\newcommand{\vslash}{v \hskip -0.5em /}

\newcommand{\bD}{{\bf D}}
\newcommand{\bDp}{{\bf D}^{(\pi)}}

\newcommand{\nn}{\nonumber \\}

\newcommand{\A}{{\cal A}}

\newcommand{\ie}{{\it i.e.}\ }
\newcommand{\eg}{{\it e.g.}\ }

\newcommand{\xipl}{\vec{\xi}_\lambda^+}
\newcommand{\ximl}{\vec{\xi}_\lambda^-}

\newcommand{\taupi}{\vec{\tau}\hskip-0.3mm
\cdot\hskip-0.3mm\vec{\pi}}

\begin{document}
\title{HADRON STRUCTURE  FUNCTIONS IN A CHIRAL QUARK 
MODEL}
\authori{Leonard Gamberg\footnote{Talk presented 
by L. Gamberg at ``Symmetry and Spin'' PRAHA-SPIN99, 
Sept. 6-11, 1999.}}
\addressi{Department of Physics and Astronomy, 
University of Oklahoma, 440 West Brooks, Norman, OK 73019}     
\authorii{Herbert Weigel\footnote{Heisenberg--Fellow}}
\addressii{Center for Theoretical Physics, Laboratory of Nuclear Science,
Massachusetts Institute of Technology, Cambridge, MA 02139}    
\authoriii{}    
\addressiii{}   
\headtitle{ HADRON STRUCTURE  FUNCTIONS IN A CHIRAL QUARK MODEL \ldots} 
\headauthor{L. Gamberg and H. Weigel}           
\specialhead{L. Gamberg and H. Weigel} 
\evidence{A}
\daterec{OKHEP-99-06, MIT-CTP-2913}    
\cislo{0}  \year{1999}
\setcounter{page}{1}
\pagesfromto{000--000}
\maketitle

\begin{abstract}
We present a consistent regularization procedure for calculating 
hadron structure functions in a bosonized chiral quark model.  
We find that the Pauli--Villars regularization scheme is most suitable. 
We also summarize the phenomenology of structure functions
calculated in the valence quark approximation.
\end{abstract}

\section{Introduction}
The strong interaction contributions to the cross--sections in deep inelastic 
scattering are parameterized by the hadronic tensor $W_{\mu\nu}$. The 
associated form factors depend on the Lorentz--scalars $Q^2=-q^2$
and $x=Q^2/2p\cdot q$, with $p$ and $q$ being the momenta of the nucleon 
and the exchanged photon, respectively. The most convincing evidence 
for the quark--substructure of the nucleon stems from the so--called 
scaling behavior of these form factors implying that certain combinations 
thereof {\it i.e.} the structure functions, depend on $Q^2$ only 
logarithmically. Without bound--state wave functions in continuum 
QCD it has proven difficult to calculate the hadron structure functions
from first principles. Fortunately the operator product expansion 
supplies an avenue to calculate nucleon structure functions: To leading 
twist (expansion in $1/Q^2$) the moments of the structure functions 
factorize into non--perturbative and perturbative pieces. The former are 
tractable within effective low energy theories for QCD while the latter 
represent the logarithmic scaling violations which can be computed 
utilizing the renormalization group equations~\cite{Ja80}. Alternatively 
one may first sum the series \ie calculate the leading twist structure 
functions at the ``model scale'', $Q^2_0$, and subsequently incorporate 
the radiative correction using the DGLAP evolution procedure.   

In a field theory calculation the most suitable starting point to 
compute hadron structure functions is the forward Compton amplitude for 
hadron photon scattering
\be
T_{\mu \nu}(p,q;s) = i
\int d^4\xi e^{iq\cdot \xi}\,
\Big\langle p,s\Big| T\left\{J_{\mu}(\xi)
J_{\nu}^{\dagger}(0)\right\}\Big| p,s \Big\rangle\, ,
\label{Comp}
\ee
since the time--ordered correlation function of electro--magnetic
currents, $J_{\mu}$ can be calculated in the functional integral formalism. 
In turn the hadronic tensor is obtained from its absorptive contribution,
\be
W_{\mu \nu}(p,q;s)=\frac{1}{2\pi}\, {\rm Im}\,
T_{\mu \nu}(p,q;s)\, .
\label{hadten}
\ee
The structure functions are then extracted from $W_{\mu\nu}$ by 
adopting the Bjorken limit, $Q^2\to\infty$ with $x$ fixed.

\section{Generalities on the Structure Functions
in the NJL--Model}

The results and discussions presented in this section are based on 
ref.~\cite{We99}. The interaction in the Nambu--Jona--Lasinio (NJL) 
model is a chirally invariant four fermion coupling, with coupling 
constant $G$. Employing functional integration techniques we may recast 
it into an effective meson theory~\cite{Eb86}. What remains is to properly 
regularize the UV divergence of the resulting determinant of the Dirac 
operator in the background of the meson fields. We utilize the 
Pauli--Villars (PV) regularization scheme which provides a formulation of 
the bosonized NJL model completely in Minkowski space~\cite{Ra95}. 
This is suitable when applying Cutkosky's rules to extract the hadronic 
tensor via~(\ref{hadten}).  In addition the PV scheme yields the proper 
scaling of the structure functions since the UV cut--off is additive in 
both the loop momenta and the (constituent) quark masses.

In the NJL chiral soliton model regularizing the nucleon structure 
functions has only recently been addressed in a satisfactory 
manner~\cite{We99}.  In the first calculations~\cite{We96,We97,Sc98}, 
contributions from the polarized vacuum to the structure functions were 
approximated within the (constituent) valence quark approximation. This 
has been motivated by the observation that the regularized vacuum 
contributions to static nucleon properties are small once the 
self-consistent soliton solution is determined. Later approaches
including vacuum contributions imposed a single subtraction regularization 
scheme at the level of (constituent) quark distributions ~\cite{Di96,Wa98}. 
The regularization attributed to a given structure function was 
conjectured to equal that of the corresponding sum rule associated with 
static nucleon properties. This approach is obviously inadequate for 
structure functions whose sum rules are not associated with a matrix 
element of a local operator. Also, the single subtraction scheme leaves
the gap--equation determining the VEV, $m=\langle S \rangle$ undefined.

To calculate the electro--magnetic current operator appearing
in the Compton tensor~(\ref{Comp}) we introduce sources $v_\mu$ which 
are conjugate to this current 
\small
\be
i \bD &=& i\dslash - \left(S+i\gamma_5P\right)
+{\cal Q}\vslash =:i\bDp+{\cal Q}\vslash
\label{defd} \\
i \bD_5 &=& - i\dslash - \left(S-i\gamma_5P\right)
-{\cal Q}\vslash =:i\bDp_5-{\cal Q}\vslash\, ,
\label{defd5}
\ee
\normalsize
with ${\cal Q}={\rm diag}(2/3,1/3)$ denoting the quark charge matrix. 
$S$ and $P$ are respectively  scalar and pseudo-scalar matrix valued 
fields. In addition to the common Dirac operator $\bD$, 
is the $\gamma_5$--odd Dirac operator $\bD_5$,
which in Euclidean space would be associated to the Hermitian
conjugate of $\bD$. The bosonized and PV regularized 
form of the NJL--model then reads 
{\small 
$\big(\sum_i c_i=\sum_i c_i\Lambda^2_i=0, \Lambda_0=0, c_0=1\big)$},
\small
\be
{\cal A}_{\rm NJL}[v_\mu]&=&
\frac{1}{4G}\int d^4x\,
{\rm tr}\left[S^2+P^2+2m_0S\right]
\label{act} \\* &&\hspace{-1cm}
-i\frac{N_C}{2}
\sum_{i=0}^2 c_i {\rm Tr}\, {\rm log}
\left[- \bD \bD_5 +\Lambda_i^2-i\epsilon\right]
-i\frac{N_C}{2}
{\rm Tr}\, {\rm log}
\left[-\bD \left(\bD_5\right)^{-1}-i\epsilon\right]\, .
\nonumber 
\ee
\normalsize
The fact that the last part of~(\ref{act}) is not 
regularized ensures that the ABJ--anomaly is completely recovered. 
The time ordered matrix element of the currents is obtained by
expanding the action to quadratic order in the sources $v_\mu$. 
In ref.~\cite{We99} we have shown that considerable 
simplifications arise in the Bjorken limit; in particular
the propagator between two insertions of the sources 
becomes that of a free quark, yielding the action
\small
\be
\A^{(2,v)}\hspace{-.35cm}&=&\hspace{-.35cm}
\frac{iN_C}{4}\sum_{i=0}^2c_i
{\rm Tr}\,\left\{\left(\bDp\bDp_5-\Lambda_i^2\right)^{-1}
\hspace{-.05cm}\left[{\cal Q}^2\vslash\left(\dslash\right)^{-1}
\vslash\bDp_5-\bDp(\vslash\left(\dslash\right)^{-1}\vslash)_5
{\cal Q}^2\right]\right\}
\nonumber \\ && \hspace{-0.1cm}
+\frac{iN_C}{4}
{\rm Tr}\,\left\{\left(-\bDp\bDp_5\right)^{-1}
\left[{\cal Q}^2\vslash\left(\dslash\right)^{-1}\vslash\bDp_5
+\bDp(\vslash\left(\dslash\right)^{-1}\vslash)_5
{\cal Q}^2\right]\right\}\, .
\label{simple}
\ee
\normalsize
The above mentioned decomposition
of the action into regularized and unregularized 
pieces has enforced the specification
($S_{\mu\rho\nu\sigma}=g_{\mu\rho}g_{\nu\sigma}
+g_{\rho\nu}g_{\mu\sigma}-g_{\mu\nu}g_{\rho\sigma}$),
\be
\gamma_\mu\gamma_\rho\gamma_\nu
=S_{\mu\rho\nu\sigma}\gamma^\sigma
-i\epsilon_{\mu\rho\nu\sigma}\gamma^\sigma\gamma^5
\quad {\rm while} \quad
(\gamma_\mu\gamma_\rho\gamma_\nu)_5
=S_{\mu\rho\nu\sigma}\gamma^\sigma+
i\epsilon_{\mu\rho\nu\sigma}\gamma^\sigma\gamma^5\, .
\label{defsign}
\ee
The latter reduction  in~(\ref{defsign}) which is reflected in 
that of $(\vslash\left(\dslash\right)^{-1}\vslash)_5$, 
arises in order that the regularization of the structure functions 
be consistent with the corresponding sum rules\footnote{This issue 
is not unique to the PV regularization scheme; in fact  all schemes 
which regularize the sum $\log\left(\bD\right)+\log\left(\bD_5\right)$ 
but not the difference $\log\left(\bD\right)-\log\left(\bD_5\right)$
 require this treatment.}. Finally from~(\ref{simple}) we calculate 
the bilocal current correlation function  and apply Cutkosky's rules 
to obtain the absorptive part~(\ref{hadten}).

The pion fields are identified as the angular variables on
the chiral circle, $S+iP=m{\rm exp}(ig\taupi/m)$ 
with the constituent quark mass being the vacuum expectation 
value (VEV) of the  scalar field, $m=\langle S\rangle$. The quark--pion 
coupling constant is extracted from the normalization of the 
pion field yielding
\be
\frac{1}{g^2}=4N_C\frac{d}{dq^2}\int_0^1 dx
\left[q^2\Pi(q^2,x)\right]\Big|_{q^2=m_\pi^2}\, ,
\label{cppil}
\ee
\vspace{-.15cm}
where 
\vspace{-.15cm}
\be
\Pi(q^2,x)&=&-i\sum_{i=0}^2 c_i\,
\frac{d^4k}{(2\pi)^4}\,
\left[-k^2-x(1-x)q^2+m^2+\Lambda_i^2-i\epsilon\right]^{-2}
\label{specfct}
\ee
is the  pion-polarization function parameterizing the quark loop.
Upon expanding $\bDp$ and $\bDp_5$ up to quadratic order in the 
pion fields we obtain the virtual Compton amplitude for pion--photon 
scattering and extract the pion structure function to be
\be
F(x)=\frac{5}{9} (4N_C g^2)
\frac{d}{dq^2}\left[q^2\Pi(q^2,x)\right]\Bigg|_{q^2=m_\pi^2}\, .
\label{pistrfct}
\ee
In the chiral limit, $m_\pi=0$, the pion structure function turns 
out to be independent of $x$ and simply reproduces the pion--quark
coupling~(\ref{cppil}). Although this result has been known~\cite{Da95} 
it serves to verify the approach defined by the 
expansion~(\ref{simple}).

In this model the nucleon emerges as a chiral soliton~\cite{Al96}.
The starting point for constructing the static soliton is the 
Dirac Hamiltonian, $h$ is defined from~(\ref{defd}),
\be
i\bDp=\beta(i\partial_t-h) \quad {\rm and}\quad
i\bDp_5=(-i\partial_t-h)\beta\, .
\label{defh}
\ee
It is diagonalized for the well--known hedgehog configuration of the 
background meson fields, 
\be
h\Psi_\alpha = \epsilon_\alpha \Psi_\alpha \qquad
{\rm with}\qquad
h=\vec{\alpha}\cdot\vec{p}+m\,\beta\, {\rm exp}
\left(i \vec{\hat{r}}\cdot\vec{\tau}\,
\gamma_5 \Theta(r)\right)\, ,
\label{diagh}
\ee
yielding eigen--spinors $\Psi_\alpha$ and energy eigenvalues
$\epsilon_\alpha$. For this mesonic configuration the action 
functional is expressed as a regularized sum over the
eigenvalues~$\epsilon_\alpha$. This yields an energy functional, 
$E\mbox{\small$[\Theta]$}$ 
which equals, to leading order in $1/N_C$, the nucleon  mass. 
The minimization of $E\mbox{\small{$[\Theta]$}}$ 
determines the localized profile function $\Theta(r)$, self--consistently. 
Nucleon states are generated by introducing collective coordinates for the
position of the soliton as well as its orientation in coordinate and flavor 
spaces. Canonical quantization of these coordinates yields the 
nucleon states which enter the matrix elements~(\ref{Comp}). Finally the 
structure functions are obtained by contracting the corresponding hadronic
tensor~(\ref{hadten}) with pertinent projectors. 
As an example we list the resulting expression for the 
polarized structure function $g_1$,
\be
g_1(x)\hspace{-0.21cm}
&=&\hspace{-0.21cm} -i\frac{M_N\mbox{\small{$[\Theta]$}}N_C}{36} 
\int \frac{d\omega}{2\pi} \sum_\alpha \int d^3\xi 
\int \frac{d\lambda}{2\pi}\, 
{\rm e}^{iM_N\mbox{\small{$[\Theta]$}}x\lambda}
\label{g1x} \\*&&\times\Bigg\{
\hspace{.25cm}
\left(\sum_{i=0}^2c_i\frac{\omega+\epsilon_\alpha}
{\omega^2-\epsilon_\alpha^2-\Lambda_i^2+i\epsilon}\right)_{\rm p}
\Big\langle N\Big| I_3 \Big| N\Big\rangle
\nn
&&\hspace{.25cm}\times
\left[\Psi^\dagger_\alpha(\vec{\xi})\tau_3
\left(1-\alpha_3\right)\gamma_5
\Psi_\alpha(\xipl)
{\rm e}^{-i\omega\lambda}
+\Psi^\dagger_\alpha(\vec{\xi})\tau_3
\left(1-\alpha_3\right)\gamma_5
\Psi_\alpha(\ximl)
{\rm e}^{i\omega\lambda}\right]
\nn
&&\hspace{.75cm}-\frac{30}{\alpha^2\mbox{\small{$[\Theta]$}}}
\Bigg[
\frac{i\lambda}{4}
\left(\frac{\omega+\epsilon_\alpha}
{\omega^2-\epsilon_\alpha^2+i\epsilon}\right)_{\rm p}
\nn
&&\hspace{.25cm}\times\Big[\Psi^\dagger_\alpha(\vec{\xi})\tau_3
\left(\hspace{-0.5mm}1-\hspace{-0.5mm}\alpha_3\right)\gamma_5
\Psi_\alpha(\xipl) {\rm e}^{-i\omega\lambda}
-\Psi^\dagger_\alpha(\vec{\xi})\tau_3
\left(\hspace{-0.5mm}1-\hspace{-0.5mm}\alpha_3\right)\gamma_5
\Psi_\alpha(\ximl){\rm e}^{i\omega\lambda}\Big]
\nn 
&&\hspace{.75cm}+\sum_\beta\left(\frac{(\omega+\epsilon_\alpha)
(\omega+\epsilon_\beta)}
{(\omega^2-\epsilon_\alpha^2+i\epsilon)
(\omega^2-\epsilon_\beta^2+i\epsilon)}\right)_{\rm p}
\langle \alpha|\tau_3|\beta\rangle
\nonumber \\ &&\hspace{.25cm} \times
\left[\Psi^\dagger_\beta(\vec{\xi})
\left(\hspace{-0.5mm}1-\hspace{-0.5mm}\alpha_3\right)\gamma_5
\Psi_\alpha(\xipl) {\rm e}^{-i\omega\lambda}
+\Psi^\dagger_\beta(\vec{\xi})
\left(\hspace{-0.5mm}1-\hspace{-0.5mm}\alpha_3\right)\gamma_5
\Psi_\alpha(\ximl){\rm e}^{i\omega\lambda}\right]
\Bigg\}\, ,
\nonumber 
\ee
where $\vec{\xi}_\lambda^\pm=\vec{\xi}\pm\lambda\hat{e}_3$ label
shifted coordinates. The subscript `p' indicates that the spectral integral
is restricted to those values of $\omega$ which cause
the respective denominators of the spectral functions to vanish.
Similar expressions are obtained for $f_1$ and 
$g_2$~\cite{We99}. For the polarized structure functions is important 
to remark that the isoscalar contribution, which is next to leading 
order in $1/N_C$ does not undergo regularization while the 
leading order isovector contribution does. 
As in the study of static nucleon properties the NJL model
we find that either the isoscalar or isovector contributions to
any structure function become regularized, but not both.  The result
that the isovector part is regularized was anticipated because
we know how to relate the zeroth moments to nucleon charges~\cite{We99}.
With regard to the unpolarized structure function $f_1(x)$ we find the 
reverse situation: a regularized leading order isoscalar and an 
unregularized next to leading order isovector contribution.
The latter piece enters the Gottfried sum rule
implying that it  is unregularized\footnote{This is in contrast
to previous studies~\cite{Di96,Wa98} where this sum rule was regularized
based upon  the assumption that it behaved equivalently to the Adler sum 
under regularization.}.

The verification of the sum rules, which relate moments of the
structure functions to static properties of hadrons, is lengthy but 
straightforward. Essentially it can be traced back to 
the expansion~(\ref{simple}) which can be formally interpreted
as a first order expansion in the operator
${\cal Q}^2(\vslash\left(\dslash\right)^{-1}\vslash)_5$. 
Such first order expansions are related to static properties.

\section{Numerical Results for the Nucleon Structure Functions}

Numerical results are still unavailable for the structure functions 
\eg~(\ref{g1x}) as projected from the consistently regularized hadronic 
tensor. Thus, we confine ourselves to presenting results from 
the valence quark approximation~\cite{We96,We97}. To compare the model 
structure functions with data, we calculate at the intrinsic scale of 
the model in the nucleon's rest frame (RF), transform them to the 
infinite momentum frame (IMF) and evolve the results to scales 
commensurate with the data. We note that the Bjorken limit singles 
out the IMF which is preferred for structure functions calculated from 
extended field configurations such as the chiral soliton because in this 
frame this class of structure functions possesses proper support~\cite{Ga98a}. 
The DGLAP evolution determines this scale parameter $Q_0^2$ at which the 
model approximates QCD.

The unpolarized structure functions are calculated from the symmetric 
portion of the hadronic tensor, $W_{\mu\nu}+W_{\nu\mu}$. In Fig. 1 we 
display the result of the structure functions entering the Gottfried 
sum rule for $e-N$ scattering as calculated in the RF and in the IMF. 
The DGLAP evolution determines our scale to be, 
$Q^2_0\approx 0.4 {\rm GeV}^2$. The resulting structure function 
reproduces the gross features of the experimental data. In addition 
depending on the other model parameter $m$, the Gottfried sum rule, 
$S_G=\int\, dx\left(F^{\rm ep}_2-F^{\rm en}_2\right)/ x$ ranges
between $0.26-0.29$ which exhibits the desired deviation from
the naive quark model prediction of $1/\ 3$ and is in reasonable
agreement with the experimental range, $0.235\pm 0.026$~\cite{Ar94}.
\begin{figure}[t]
\centerline{
\epsfig{figure=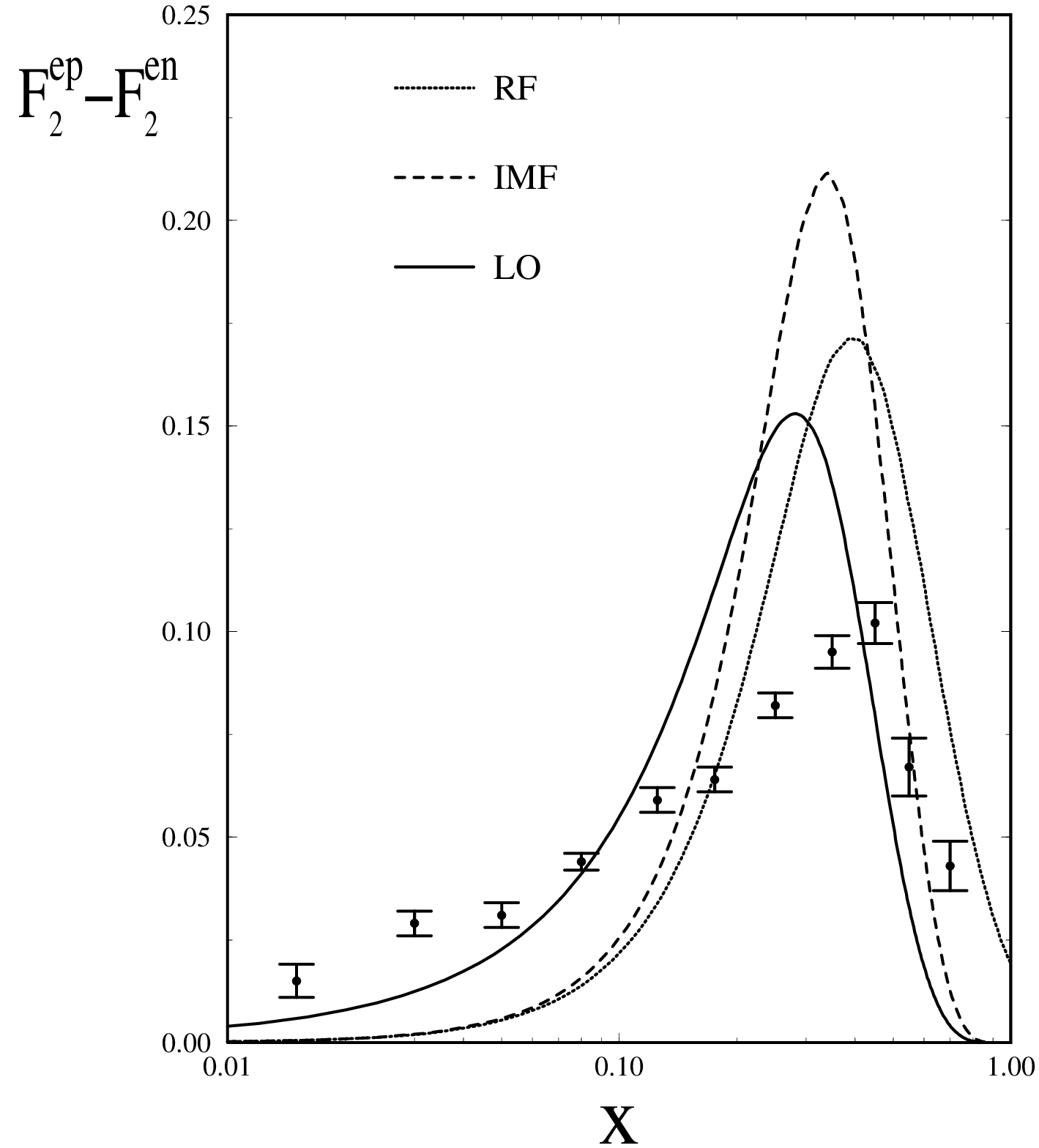,height=4.0cm,width=5.5cm}
\hspace{0.75cm}
\epsfig{figure=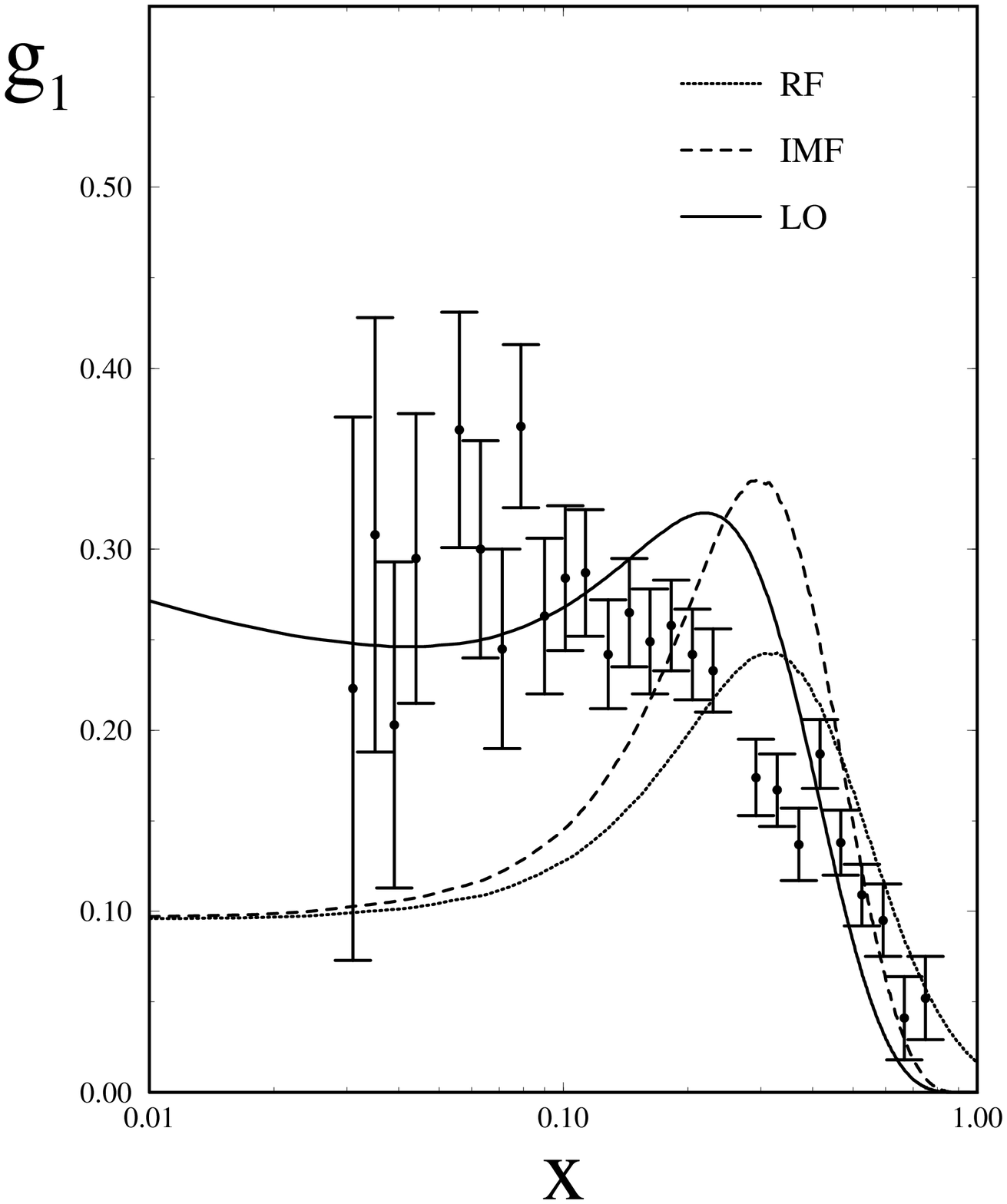,height=4.0cm,width=5.5cm}}
\caption{
{\small
Left Panel: The unpolarized structure function entering the
Gottfried sum rule: RF, IMF and leading order QCD evolution to 
$Q^2=3 {\rm GeV}^2$, with full $1/N_c$ 
contribution; see \protect\cite{Sc98}.
Right Panel: $g_1(x)$.}}
\label{sgott}
\end{figure}

The spin--polarized structure functions $g_1$ and $g_2$
are extracted from the anti-symmetric combination, $W_{\mu\nu}-
W_{\nu\mu}$. Results are shown in Fig. 2 and are compared
with the data from SLAC~\cite{slac96}. In particular
we note the sizeable twist two and three contributions.
The twist two portion is given by the
Wandzura--Wilczek  contribution, the first two
terms in
\be
g_2\left(x,Q^2\right)
=-g_1\left(x,Q^2\right)+\int_x^1\frac{dy}{y}g_1\left(x,Q^2\right) 
+ \overline{g}_2\left(x,Q^2\right)\, ,
\ee
where $\overline{g}_2$ denotes the twist three contribution.
In the evolution of the twist-2 portion we have restricted
ourselves to the leading order in $\alpha_s$  since
the next to leading order evolution of the twist--3
$\overline{g}_2\left(x,Q^2\right)$ is not known
(The twist--3 piece is known only in
the large $1/N_{\rm c}$ limit~\cite{Bal96}.).
\begin{figure}[t]
\centerline{
\epsfig{figure=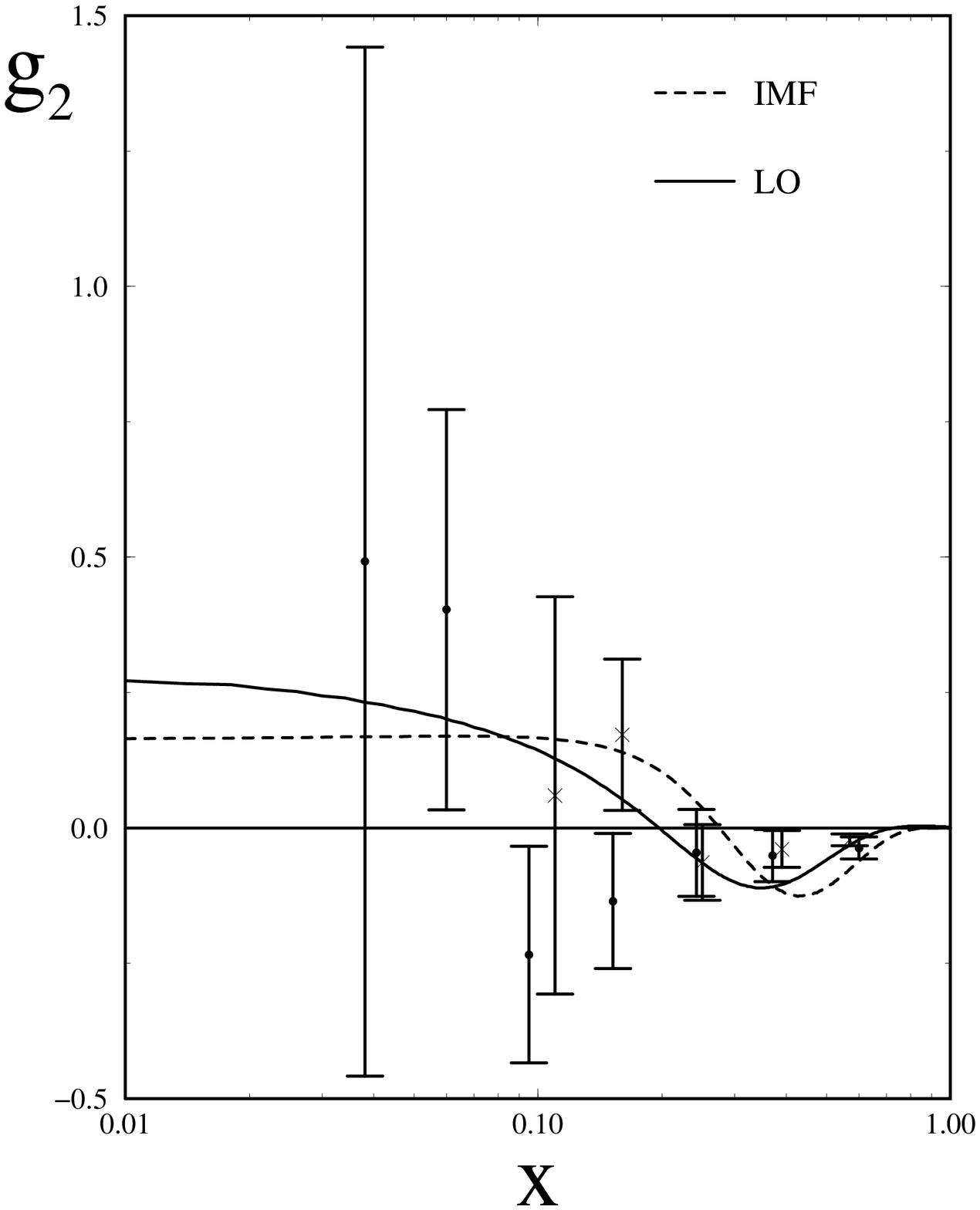,height=4.0cm,width=5.5cm}
\hspace{.75cm}
\epsfig{figure=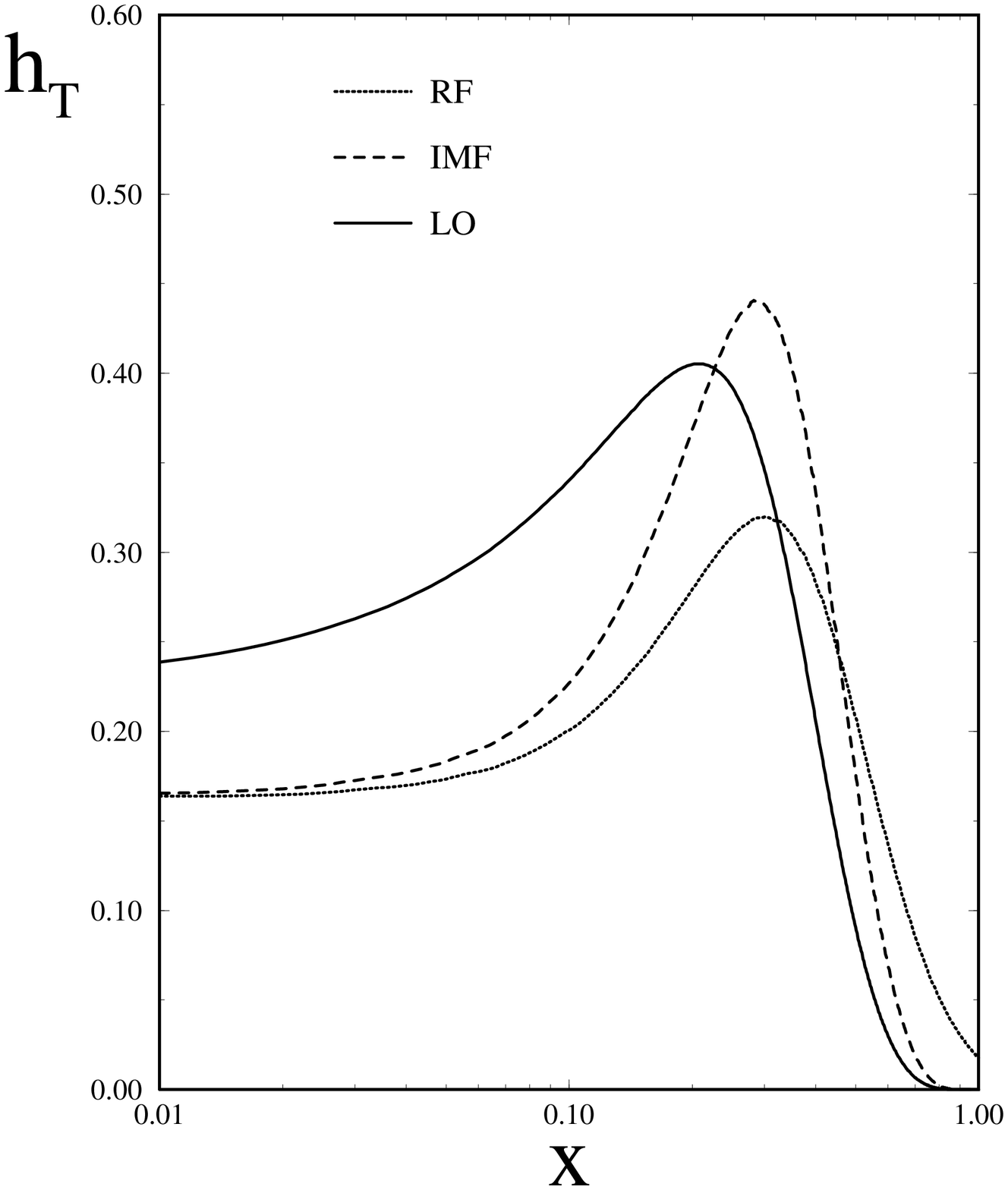,height=4.0cm,width=5.5cm}}
\vspace{-.25cm}
\caption{
{\small 
Left: $g_{2}^{p}\left(x,Q^2\right)=g_2^{WW(p)}\left(x,Q^2\right)+
{\overline{g}}_{2}^{p}\left(x,Q^2\right)$ evolved from 
$Q_0^{2}=0.4{\rm GeV}^2$ to $Q^2=5.0{\rm GeV}^2$. Right Panel: 
Chiral odd structure function $h_{T}^{p}\left(x,Q^2\right)$.}}
\label{fig_g2}
\end{figure}
In ref.~\cite{We99} the Bjorken $\int dx(g_1^p(x)-g_1^n(x))=g_A/6$ 
and Burkhardt--Cottingham $\int dx\, g_2(x)=0$ sum rules have been 
explicitly verified.

Having calculated the spin (un)polarized structure functions, $f_1$ and 
$g_1$ and their constituent quark components we investigate the chiral 
odd flavor distributions and their charge weighted average nucleon ``structure 
functions''~\cite{Ga98b}. Here we confine ourselves to the presentation 
of results relevant in the context of the Soffer inequality~\cite{So95}. 
This inequality relates the nucleon chiral odd quark distribution functions 
to both the unpolarized $f_1^{(u)}(x,Q_0^2)$ and polarized 
$g_1^{(u)}(x,Q_0^2)$ chiral even counter--parts.
\be
f_1^{(u)}(x,Q^2)+g_1^{(u)}(x,Q^2)\ge 2\ h_T^{(u)}(x,Q^2)\ .
\label{sofineq}
\ee
The superscript refers to the flavor combination which projects 
onto up--quark quantum numbers. Note again, that this projection 
refers to the constituent quarks which contain some non--perturbative 
gluonic distributions. Fig. 3 clearly demonstrates that the inequality
is satisfied at the model scale.  
\begin{figure}[h]
\centerline{
\epsfig{figure=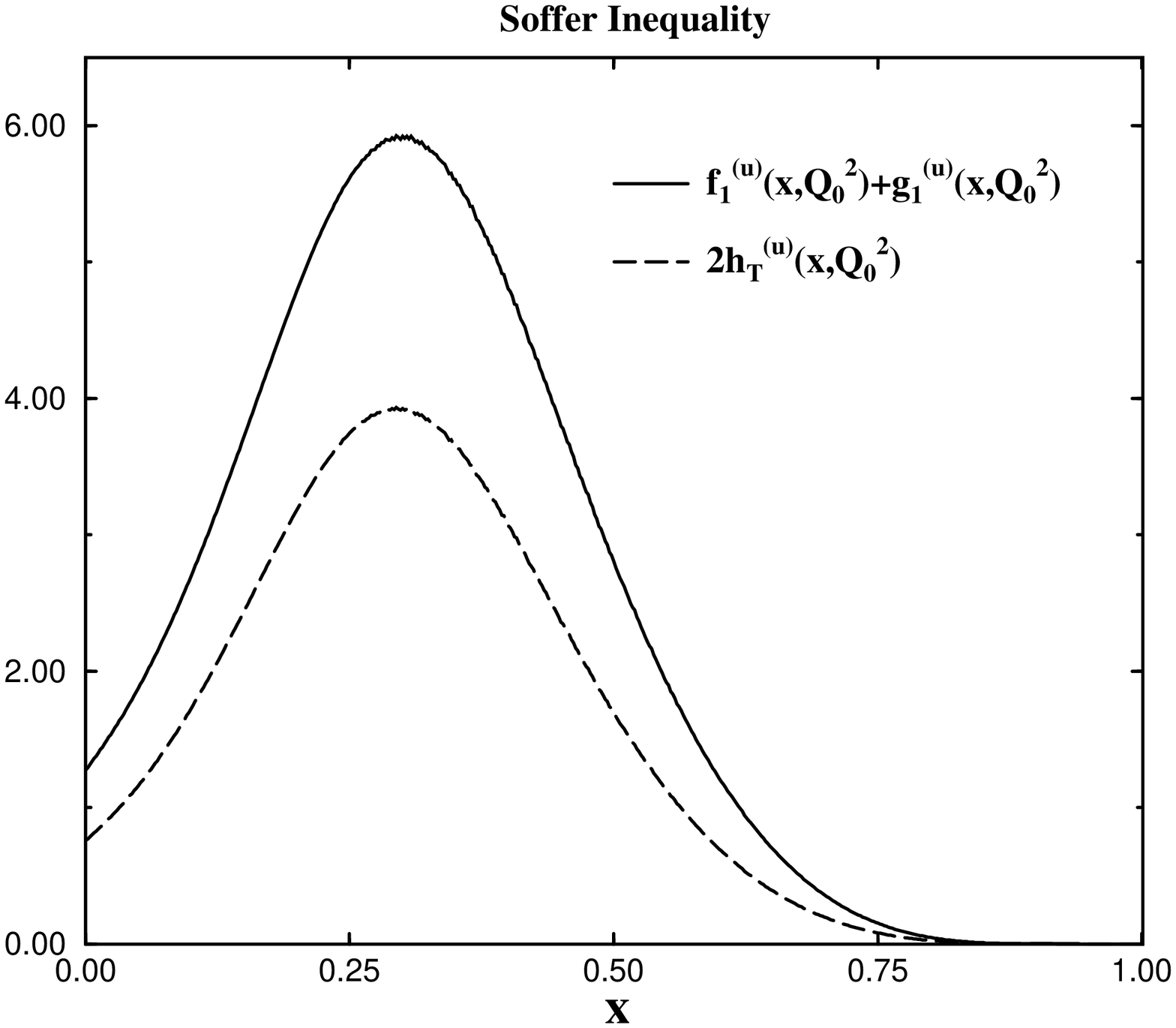,height=5.5cm,width=5.5cm,angle=270}
\hspace{.75cm}
\epsfig{figure=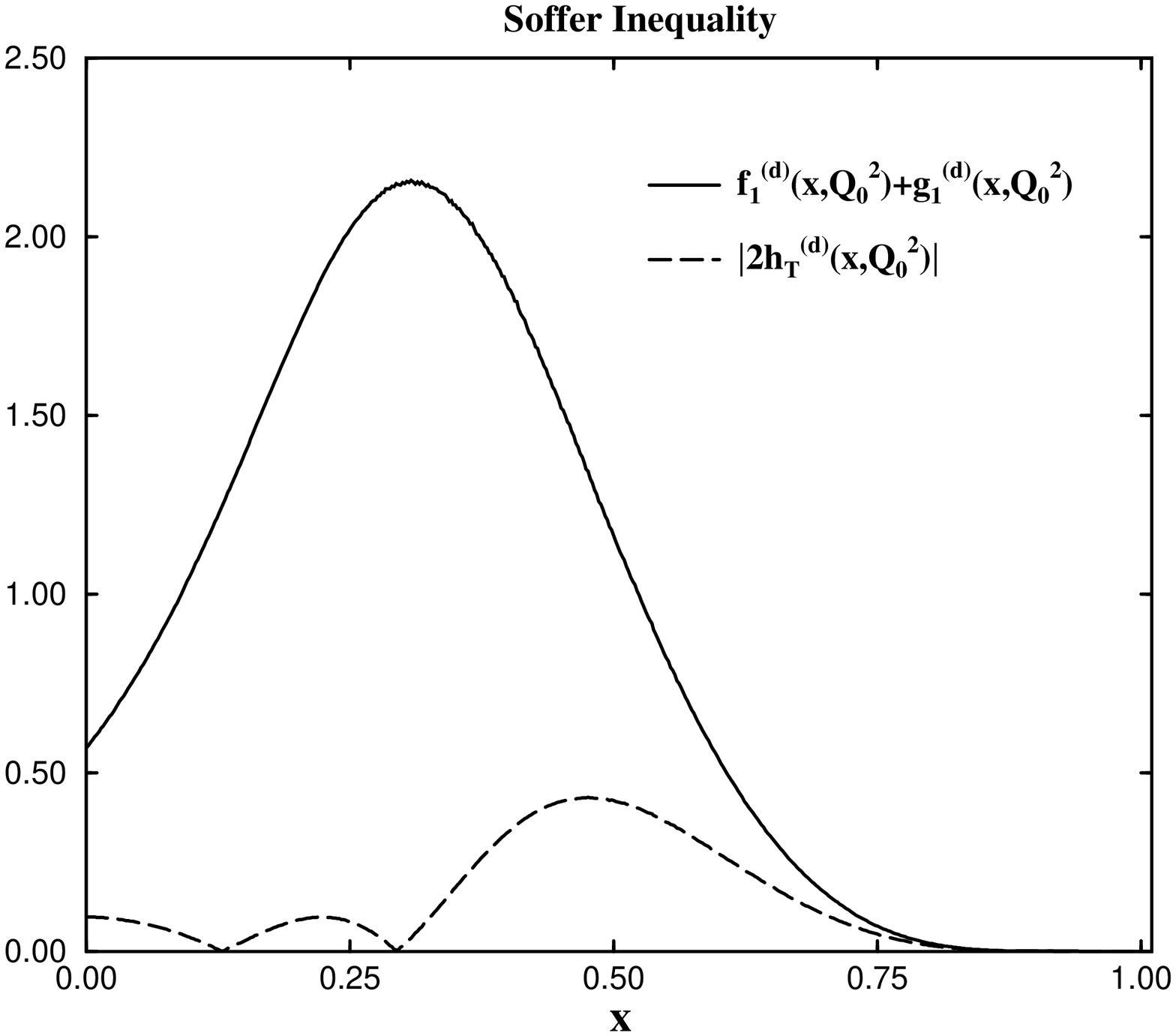,height=5.5cm,width=5.5cm,angle=270}}
\vspace{-.75cm}
\caption{
{\small
Left Panel: The Soffer inequality for the chiral even 
combination in the IMF  (solid line)
for  the up--quark distributions
and the chiral odd structure function $2h_T^{(u)}(x,Q_0^2)$
(long--dashed line). Right Panel: Down quark distributions.}}
\label{soff}
\end{figure}

\section{Conclusions}
We have summarized our results on the consistent regularization
scheme 	in calculating  hadron structure functions in the bosonized 
NJL chiral soliton model where baryons emerge as chiral solitons. 
We have calculated the Compton amplitude in the Bjorken limit from the  
PV regularized action functional with the constraint of preserving the 
anomaly structure of QCD. Introducing pertinent external sources enables 
the unambiguous identification of the quark currents which in turn 
allowed us to calculate the bilocal current correlation function entering
the forward Compton amplitude.  This scheme  satisfies the leading
order scaling properties of nucleon structure functions.
The sum rules for both polarized and unpolarized are fulfilled 
because in the associated moments Cutkosky's rules extract -- similar to
the Cauchy integrals -- the matrix elements associated with static
nucleon properties~\cite{We99}. These results represent the first step 
towards a fully consistent calculation of structure functions in the 
NJL model. We have also summarized our results in the valence quark 
approximation. Due the sizeable contribution from this state to all
nucleon properties in the NJL model, not surprisingly, this approximation 
reproduces the general trends of the data on structure functions.

\section*{Acknowledgement}
\vspace{-.20cm}
{\small
We gratefully acknowledge contributions from
our co--workers H. Reinhardt and E. Ruiz Arriola.
L.G. wishes to thank the organizers of SPIN-99 workshop for their 
efforts and hospitality and to K.A. Milton for support on 
this project.  This work is supported by funds provided by the 
U.S. Department of Energy (D.O.E.)
\#DE--FG03--98ER41066, and \#DF--FC02--94ER40818,
and the Deutsche Forschungsgemeinschaft (DFG) 
under contract We 1254/3-1.}

\end {document}